\documentclass[graybox]{svmult_gam}

\usepackage{mathptmx}       
\usepackage[varg]{txfonts}  
\usepackage{helvet}         
\usepackage{courier}        
\usepackage{type1cm}        
\usepackage{makeidx}         
\usepackage{graphicx}        
\usepackage{multicol}        
\usepackage[bottom]{footmisc}
\usepackage{natbib}
\usepackage{journals}
\setcitestyle{numbers,square}

\makeindex             

\begin{document}

\title*{Predicting the frequencies of young and of tiny galaxies}
\author{G. A. Mamon, D. Tweed, T. X. Thuan and A. Cattaneo}
\institute{G. A. Mamon \at IAP (UMR 7095: CNRS \& UPMC), Paris, France, \email{gam@iap.fr}
\and D. Tweed \at Racah Instability. of Physics, Hebrew University,
Jersusalem, Israel
\and T. X. Thuan \at Dept. of Astronomy, University of Virginia,
Charlottesville, VA, USA
\and A. Cattaneo \at Laboratoire d'Astrophysique de Marseille, France}
%
%

\maketitle

\abstract{A simple, one-equation, galaxy formation model is applied to both
the
  halo merger tree derived from a high-resolution
dissipationless cosmological simulation and to a quarter million Monte-Carlo 
halo merger
trees.
The galaxy formation model involves a sharp entropy barrier against the accretion
of gas onto low-mass halos, the shock heating of infalling gas far from the
central regions of massive halos, and supernova feedback that drives the gas out of shallow
halo potential wells.
With the first approach, we show that the large majority of galaxies within
group- and cluster-mass halos, known to be mainly dwarf ellipticals,  
have acquired the bulk of their stellar 
mass through gas
accretion and not via galaxy mergers.
With the second approach, 
we qualitatively reproduce
 the downsizing trend of greater ages at greater masses in stars and
predict an upsizing trend of greater ages as one proceeds to masses lower
than $10^{10} M_\odot$.
We find that the fraction of galaxies with very young stellar populations
(more than half the stellar mass formed within the last 1.5 Gyr) is a
function of present-day stellar mass, which peaks at 0.5\% at
$m_{\rm crit}$=$10^{7.5-9.5} M_\odot$, roughly corresponding to the masses of
blue compact dwarfs.
We predict that the stellar mass function of galaxies should not show a
maximum at $m_{\rm stars} > 10^{5.5}\, M_\odot$, with a power-law stellar
mass function with slope $\approx -1.6$ if the IGM temperature in the
outskirts of halos before reionization is set by molecular Hydrogen cooling.
We
speculate on the nature of the lowest mass galaxies.
}

\index{galaxy formation}
\index{reionization}
\index{ages}
\index{downsizing}
\index{upsizing}
\index{star formation efficiency}
\index{feedback}
\index{gas accretion}
\index{mergers}
\index{young stellar populations}
\index{merger trees}
\index{mass function}
\index{I~Zw~18}
\index{blue compact dwarfs}
\index{dwarf spheroidals}
\index{ultra-faint dwarfs}
\index{globular clusters}
\index{high-velocity clouds}

\section{Introduction}
The mass growth of galaxies can occur 
either by 
 accretion of gas that cools to form molecular clouds in which stars form
or
by galaxy mergers.
While spiral disks are believed to form through the first mode,
it is still unclear whether elliptical galaxies are built by mergers or not.
We use a very simple \emph{toy} model of galaxy formation \citep{CMWK10} run
on top of the dark
matter halo merger tree obtained from a high-resolution dissipationless
cosmological simulation (CS) to understand how dwarf galaxies
acquire their mass. We also use our model to predict the frequency of
galaxies
such as the very metal-poor (1/50th solar
metallicity) galaxy I~Zw~18, for which the bulk of the stellar mass is
younger than $<1$ Gyr (\citep{IT04,TYI10}, see also \citep{Aloisi+07} using
HST color-magnitude diagrams) or 500 Myr  (\citep{Papaderos+02,HTI03} from
photometric studies).

\section{Galaxy formation model}

Galaxies form in DM halos, and 
our
toy model 
gives the mass in stars and cold gas, $m$, as a function of halo
mass $M$ and epoch $z$, taking into account the fact that for stars to form
one needs: 1) gas accretion, which is fully quenched for low-mass halos
\citep{TW96,Gnedin00}; 2) in cold form, which becomes inefficient in
high-mass halos \citep{BD03,Keres+09a}; and 3) to retain the interstellar
gas against supernova (SN) winds \citep{DS86}:
\begin{equation}
m_{\rm stars}(M,z)={v_{\rm circ}^2-v_{\rm reion}^2 \over v_{\rm circ}^2/(1-g) + v_{\rm
    SN}^2}\,{f_{\rm b}\,M \over 1 + M/M_{\rm shock}} \ ,
\label{toy}
\end{equation}
where $f_{\rm b}$=$\Omega_{\rm b}/\Omega_{\rm m}\simeq 0.17$ 
is the cosmic baryon fraction,
$v_{\rm reion}$ is
the minimum halo circular velocity for gas accretion (which rises abruptly
after reionization; $m_{\rm stars} =0$ for $v_{\rm circ} < v_{\rm reion}$),
$v_{\rm SN}$ is a characteristic velocity for
SN feedback, 
$M_{\rm shock}$ represents the transition from 
pure
cold 
to mainly hot accretion, and $g$ is the fraction of retained baryons in the
form of cold gas.

%
\begin{figure}
\centering
\includegraphics[width=7.1cm]{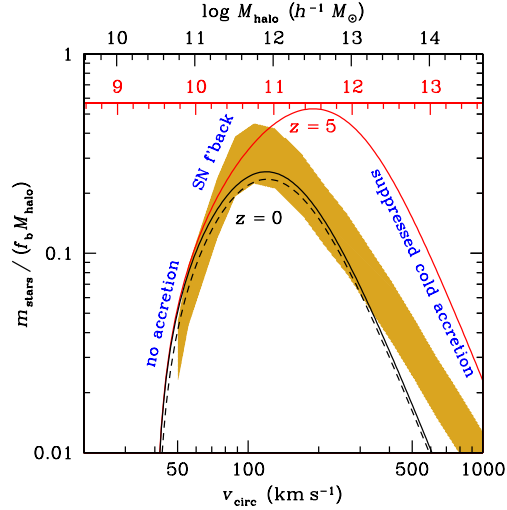}
\caption{Illustration \citep{CMWK10} of the toy model of galaxy formation
  (eq.~[\ref{toy}] with $g=0$, $v_{\rm reion}$=$40 \, \rm km \, s^{-1}$, $v_{\rm
    SN}$=$120 \, \rm km \, s^{-1}$, $M_{\rm shock}$=$8\times
  10^{11}\,h^{-1} M_\odot$ 
  and 40\% of the stellar mass stripped at every passage through a parent
  halo)  at  $z$=$0$ (\emph{upper halo mass scale}) and $z$=$5$ (\emph{lower halo mass
  scale}). 
The variations with redshift come from the redshift modulation of
  the dependence of $v_{\rm circ}$ on $M$.
The \emph{dashed curve} shows the effect of our iterative correction for
  $g$ (\S\,\ref{when}).
The \emph{shaded region} comes from abundance matching \citep{GWLBK10}.
The disagreement, at the
high end,  of our model with the abundance matching prediction suggests that
galaxy mergers are
important at high halo masses (see \citep{CMWK10}).
\label{sfe}}
\end{figure}
Figure~\ref{sfe} describes the efficiency of galaxy formation, $m_{\rm
  stars}/(f_{\rm b} M)$, at $z$=0 and 5, with the parameters tuned to 
  match the observed ($z$$\simeq$0.1)
galaxy 
stellar mass function (MF) of \citep{BMcIKW03}. 
Galaxy formation 
occurs in a fairly narrow range of halo masses, that varies with redshift.
\begin{figure}
\centering
\includegraphics[width=9cm]{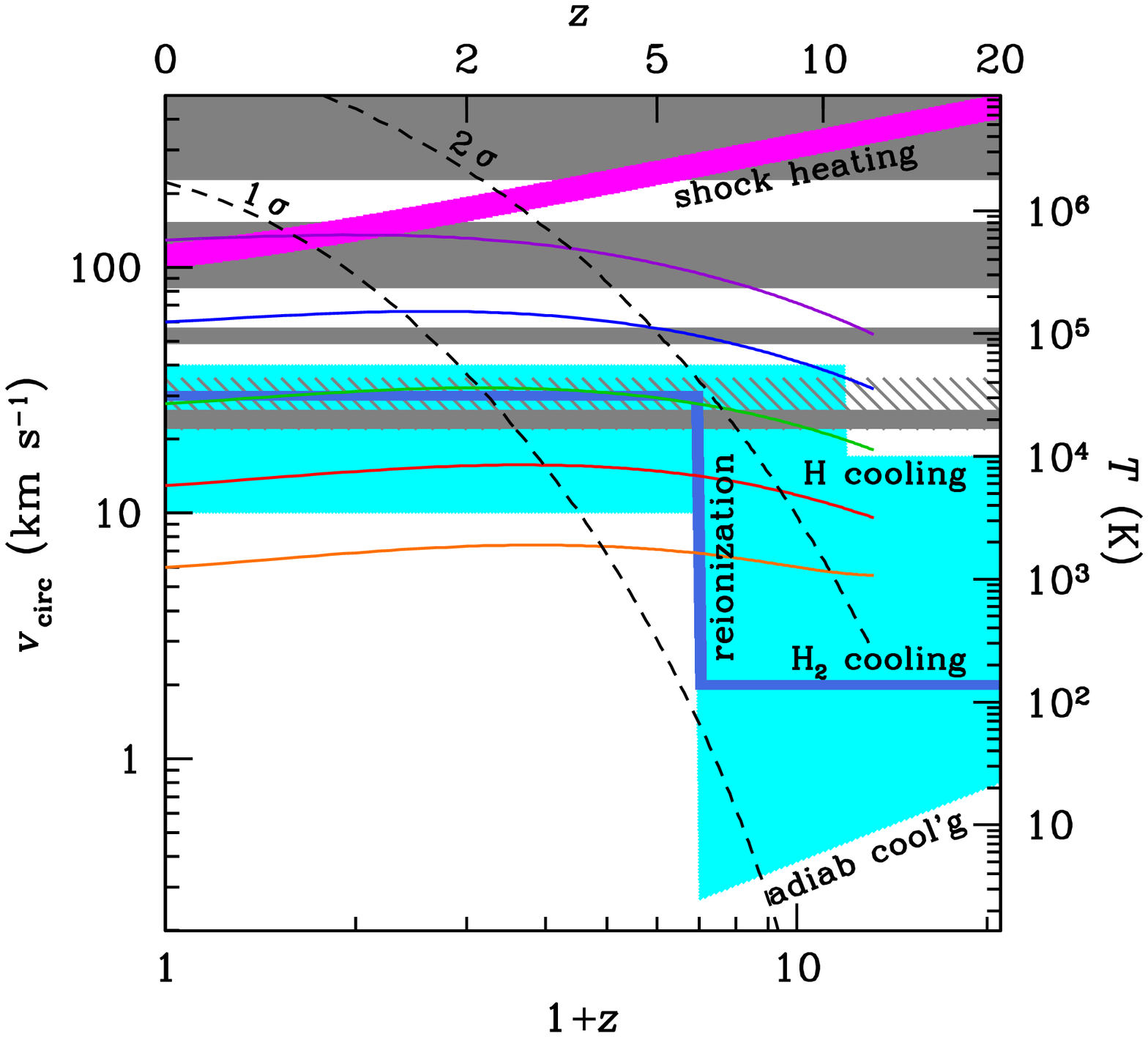}
\caption{Halo circular velocity versus redshift. 
Logarithmic mean evolution (\emph{solid curves});
typical 1 and $2\,\sigma$ fluctuations in $\varLambda$CDM scenario 
(\emph{dashed curves}).
The \emph{thick broken solid line} indicates our adopted history of the
temperature of the IGM (just outside the virial radius of galaxies), while
the \emph{cyan hashed region around it} shows the uncertainty on this
temperature. The equivalent circular velocity is the minimum for gas
accretion
onto halos.
The \emph{thick rising magenta curve} is the limiting circular velocity for shock
heating of infalling gas near the virial radius.
The \emph{gray horizontal bands} indicate temperatures where solar and
subsolar (\emph{narrow hashed}) and also 1\% solar (\emph{wide hashed}) metallicity gas is
thermally unstable.
\label{vvsz}
 }
\end{figure}
This is further illustrated in Figure~\ref{vvsz}, which shows the time
evolution (from right to left) of halos ending with different circular
velocity at $z=0$. Interestingly, halos evolve,  on average,  with nearly constant
circular velocity (however individual halos have constant $M$, hence decreasing $v_{\rm circ}$ in
their quiescent mode, so this decrease is 
compensated by the
increases caused by mergers).
Also, the temperatures affected by thermal instability (TI) are fairly narrow for
the lowest mass galaxies, and TI can therefore be neglected to first order in our analysis.

We first apply equation~(\ref{toy}) to the merger tree obtained from the halos and
subhalos (AHF algorithm of \citep{KK09}) 
of a high resolution dark matter CS.
When a halo enters a
more massive one, it becomes a \emph{subhalo} and its galaxy becomes a
\emph{satellite}. The subhalo orbit is followed 
until either 1) dynamical friction (DF) causes it to fall to the halo
center and necessarily see its galaxy merge with the central galaxy of the
parent halo, or 2) it is tidally stripped and heated by the global halo potential
to the point that there are insufficient particles to follow it. In this
latter case ($M_{\rm subhalo} < 1.5$$\times$$10^9 \,h^{-1} M_\odot$,
corresponding to $v_{\rm circ}=17 \, \rm km \, s^{-1}$ at $z=0$), 
we assume that the satellite galaxy merges with the central one
after a delay set by DF, for which we adopt the
timescale carefully 
calibrated by \citep{Jiang+08} with hydrodynamical CSs. 

The present-day relation between stellar and halo mass obtained when
our
model (with $g=0$, which is adequate for massive galaxies)
is applied to the CS \citep{CMWK10} 
is impressively close to the
prediction derived by abundance matching \citep{GWLBK10}.
Moreover, the variation of the stellar mass functions in different bins of
halo mass matches very well the measurements from the SDSS \citep{YMvdB09}.

\begin{figure}
\centering
\includegraphics[angle=-90,width=7.6cm]{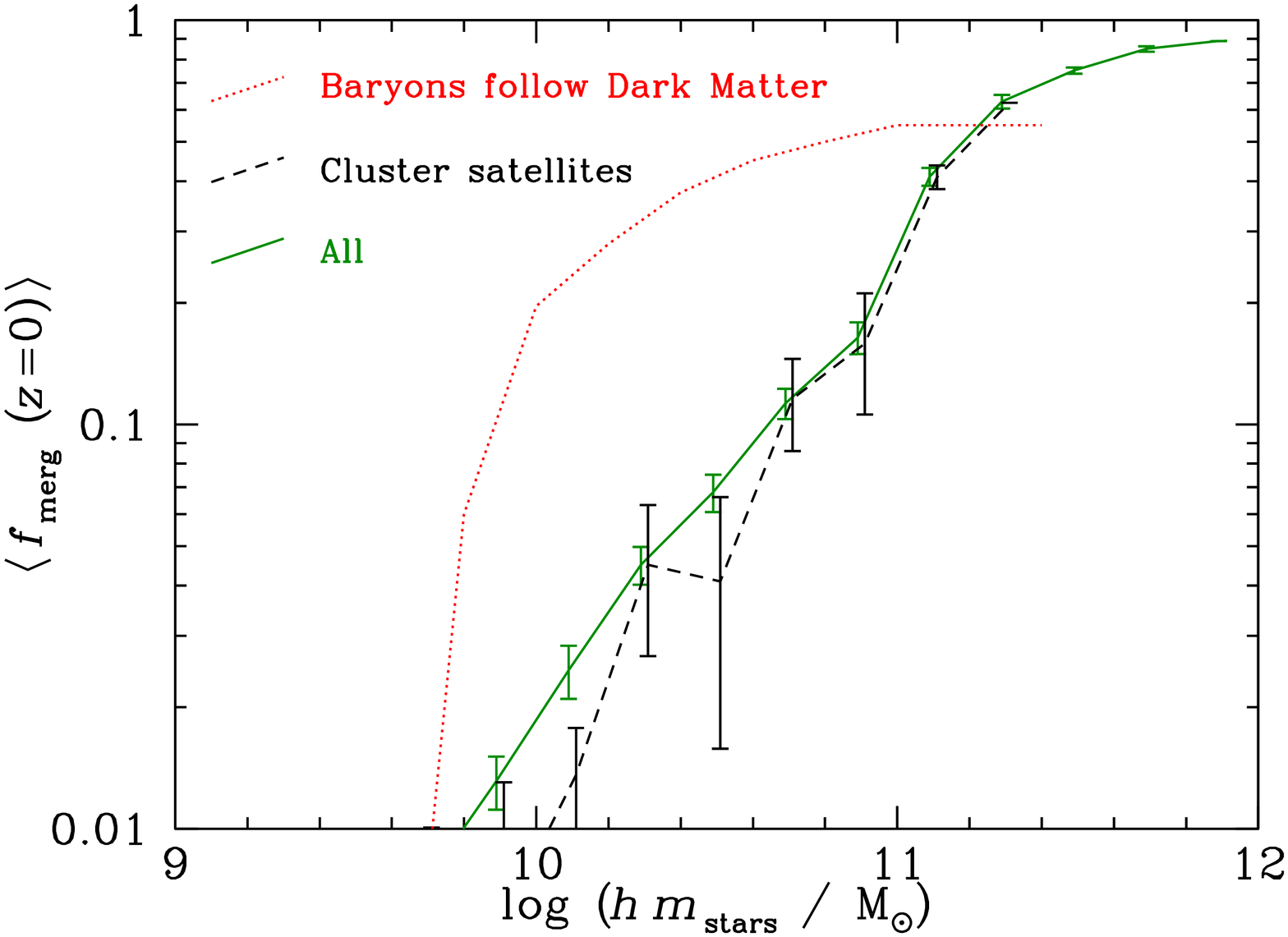}
\caption{Median fraction of $z$=0 stellar mass (for $h m_{\rm stars} >
  10^{10} M_\odot$ one can neglect the gas) acquired by mergers, for all galaxies
  (\emph{solid green}) and $h\,M_{\rm halo}> 10^{13} M_\odot$ cluster
  satellites (\emph{dashed black}). The \emph{red dotted line} shows the
  baryons trace the dark matter model.
The error bars are
  uncertainties on the median from 100 bootstraps. 
\label{fmvsm}}
\end{figure}
Figure~\ref{fmvsm} 
shows that
while mergers dominate the growth of the massive galaxies (as expected from
the toy model, since gas accretion is quenched at high masses), their
importance drops sharply when one moves to stellar masses below $10^{11}\,h^{-1}
M_\odot$ (the mass resolution is $m_{\rm stars}^{\rm min} \simeq 10^{10.6} \,h^{-1} M_\odot$,
where the median fraction of stellar mass acquired by mergers no longer decreases
with $m_{\rm stars}$ faster than in our reference model where baryons trace the dark
matter [red dotted line]).
This dominance of gas accretion at low mass is also true for the satellites
of clusters (dashed line). Since observations indicate that most satellites
of clusters are 
dwarf ellipticals (dEs), we conclude that cluster dEs are most often 
not built by mergers. One must resort to other mechanisms (not included in
our toy model) that transform dwarf irregulars into dEs
(e.g. harassment \citep{Mastropietro+05} or ram pressure stripping \citep{BBCG08}).

\section{When do dwarf galaxies form their stars?}
\label{when}
We have used the halo merger tree code of \citep{ND06} to statistically study
the star formation histories (SFHs) of dwarf galaxies. We consider 24 final
halo masses geometrically spaced between $10^7$ and $10^{12.75} \,h^{-1}
M_\odot$, and run each halo merger tree 10$\,$000 times.  We run the toy
model (eq.~[\ref{toy}] with $g=0$) on the branches of the halo merger tree (moving
forward in time) to follow the evolution of stellar mass.  To compare our
predictions to SDSS measurements (at $z\simeq0$), we correct our $z=0$
stellar masses for the non-inclusion of gas in equation~(\ref{toy}), assuming
that the gas-to-star fraction is
\begin{equation}
{\rm G \over S} = {g\over 1-g} = {\rm dex} \left (a + b\, \log m_{\rm
  stars} \right ) \ ,
\label{GoverS}
\end{equation}
where we derive $a=4.5$ and $b=-0.5$ by fitting a straight line through
Fig.~11 of \citep{BGD08}. This correction involves solving an implicit
equation, which we perform iteratively.

\begin{figure}
\centering
\includegraphics[angle=-90,width=0.9\hsize,bb=120 1 506 760]{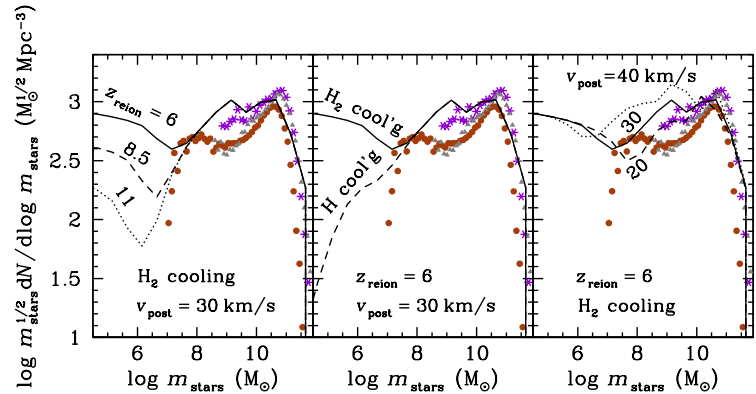}
\caption{Effects of thermal history of IGM on galaxy stellar mass function.
\emph{Left:}
Effect of reionization epoch.
\emph{Middle:}
Effect of pre-reionization IGM temperature.
\emph{Right:}
Effect of post-reionization IGM temperature.
In all plots, the \emph{symbols} 
represent the observed SDSS  \emph{stellar} mass
functions, 
measured by 
\citep{BMcIKW03} (\emph{purple asterisks}),
\citep{BGD08} (\emph{brown circles}, strongly incomplete below $\log m_{\rm
  stars}  = 7.4$), 
and \citep{YMvdB09} (\emph{gray triangles}).
\label{masfuns}}
\end{figure}

%
The predicted MFs
match fairly well the observed MFs.
The left panel of Fig.~\ref{masfuns}
suggests that reionization must occur late ($z$=6).
The middle panel 
hints that, before reionization, the temperature of the IGM
must be set by molecular cooling ($v_{\rm pre-reion}=2 \, \rm km \, s^{-1}$).
The right panel indicates a good match between predicted and observed
stellar MFs 
when the IGM after
reionization is not too cool ($v_{\rm post-reion}\geq30 \, \rm
km \, s^{-1}$).

\begin{figure}
\includegraphics[width=0.49\hsize]{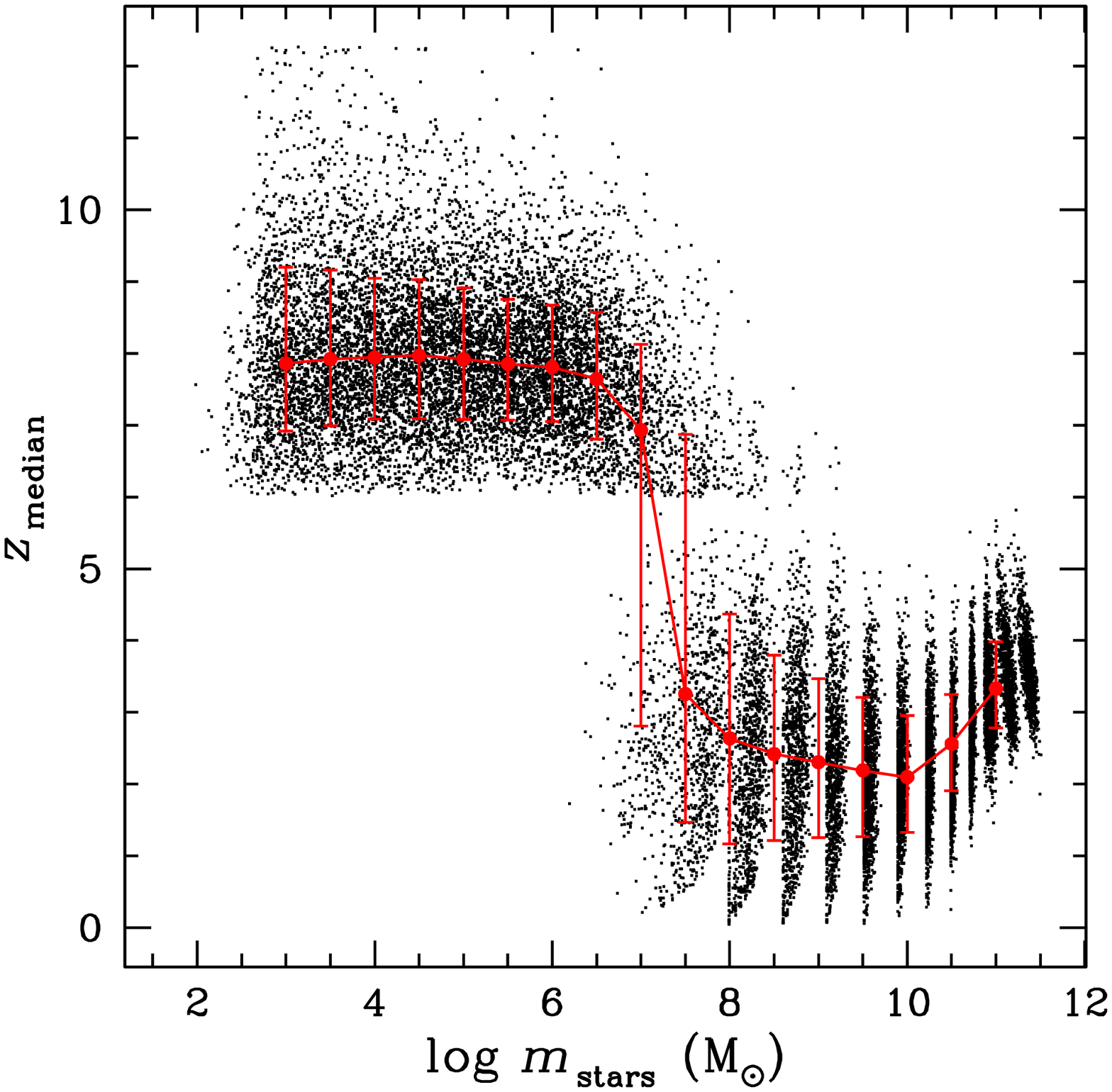}
\quad
\includegraphics[width=0.49\hsize]{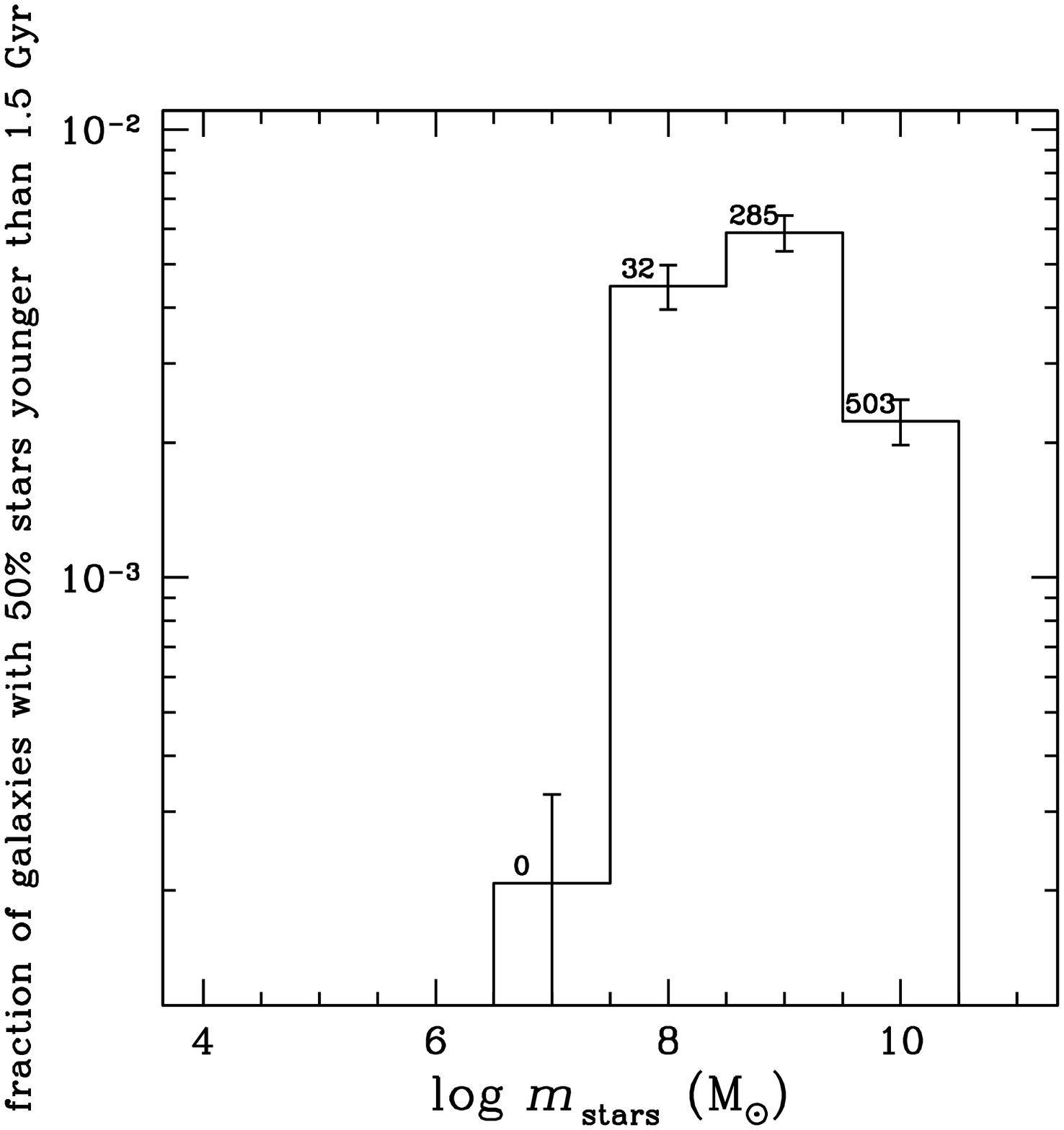}
\caption{\emph{Left:} 
Median (mass-weighted) star formation redshift vs $z$=0 mass. The points (1
in 5 plotted for clarity) are
individual galaxies, while the red symbols are medians (error bars extend from
16th to 84th percentiles). The stripes are artifacts
of our discrete set of final halo masses.
\emph{Right:}
Fraction of young galaxies (half the mass formed within 
  last 1.5 Gyr) versus $z$=0 stellar mass (with
numbers of expected SDSS galaxies).
In both plots, we adopt $v_{\rm pre-reion}$=$2 \, \rm km \, s^{-1}$ ($\rm H_2$
cooling) before
reionization ($z$=6), $v_{\rm post-reion}$=$30 \, \rm km \, s^{-1}$ for $z$$<$6,
$v_{\rm SN}$=$120 \,\rm km \, s^{-1}$, and
$h M_{\rm shock}$=$8\!\times\!10^{11} \,M_\odot$. 
}
\label{fracyoung}
\end{figure}

According to
the left panel of Figure~\ref{fracyoung},
as one
proceeds from the highest final stellar masses to lower ones,
the median stellar age first diminishes, qualitatively
reproducing the \emph{downsizing} of star formation. However this downsizing
stops at  $m_{\rm stars} \approx 10^{10} M_\odot$  and as one proceeds to even
lower masses, one notices an \emph{upsizing} of stellar ages, first weak,
becoming strong at $m_{\rm stars} <10^{7.5}M_\odot$. In our model, 
the smallest galaxies form most of their stars before reionization.
Classical dwarf spheroidals, with $m_{\rm stars} = 10^8 M_\odot$ should have
moderately old stellar populations, with a large scatter (caused by the
nearly constant average evolution of halo circular velocity), while
ultra-faint dwarfs should be extremely old (see also \citep{SF09}).
Note that our model predicts a discontinuity in median age versus 
mass at $\approx 10^7 M_\odot$, which appears to be in conflict with the
continuity of the metallicity-mass relation \citep{Lee+06}.

Young galaxies must cross the low-mass (entropy) barrier (thick broken line
in Fig.~\ref{vvsz}) only very late.
The right panel of Figure~\ref{fracyoung} shows that the frequency of
galaxies with the bulk of their mass in stars acquired within the last 1.5
Gyr is maximal at $\simeq 0.5\%$ at $m_{\rm stars}$=$10^{7.5-9.5}
M_\odot$.  The presence of a young galaxy such as I~Zw~18, whose stellar mass
is of order $10^{6.7} M_\odot$ \citep{TYI10}, is consistent with our model:
we predict 0.02\% of young galaxies at $\log m = 7\pm0.5$, and with 941 in
this mass range in the SDSS, we expect 0.2 young galaxies in SDSS at this
mass, hence (from Poisson statistics) there is a probability of 18\% of
detecting at least one galaxy as young as I~Zw~18 in this mass bin. 
 
Our model also predicts several hundred
young galaxies in the SDSS, mostly at $m_{\rm stars} = 10^{10} M_\odot$
(fraction of 0.2\%).  In comparison, a first analysis of luminosity-weighted
ages of high surface brightness SDSS galaxies by \citep{Gallazzi+05} leads to
$\approx 100$ times higher fractions of young galaxies: 16\% with ages younger
than 1 Gyr at $m=10^{10} M_\odot$.  A second analysis by the same team, this
time of mass-weighted stellar ages for a large fraction of SDSS galaxies
\citep{GBCW08}, reveals much lower fractions of young galaxies: 1.4\% at $\log
m_{\rm stars}=10\pm0.5$, but still as much as 15\% at $\log m_{\rm
  stars}=9\pm0.5$, still
respectively 
 7 and 30 times higher than our
predictions. Also, the fraction of ages (summing over all masses) less than 4
Gyr derived from a much more-refined semi-analytical model \citep{dLB07} run
on the Millennium cosmological dark matter simulation \citep{Springel+05} is
negligible in comparison with \citep{GBCW08}'s prediction of 3\%.  This
discrepancy in the predicted fractions of young stellar populations among
galaxies of intermediate mass is thus an open issue worth further
exploration.

\section{The lowest mass galaxies}

The middle panel of Figure~\ref{masfuns} indicates that there is no peak in
the best fitting galaxy MF, if the IGM
temperature before reionization is set by molecular
Hydrogen cooling ($v_{\rm pre-reion}$=$2 \, \rm km \, s^{-1}$) with a low-end
slope of $-1.59$. This is in excellent agreement with the slope of $-1.58$
found by \citep{BGD08}. If, instead, the IGM temperature before reionization is set
by atomic Hydrogen cooling ($v_{\rm pre-reion}$=$17 \, \rm km \, s^{-1}$), the mass
function peaks at $m_{\rm stars}$=$10^{5.5} \,h^{-1}
M_\odot$. This maximum is probably not caused by of our mass resolution,
since no such peak is seen when the pre-reionization IGM temperature is set by
$\rm H_2$ cooling.

The importance of the low-end tail of the galaxy MF  raises the
question of the nature of very low mass objects ($m_{\rm stars} <
10^6 M_\odot$).
Two classes of objects come to mind: Globular Clusters (GCs) and High Velocity
Clouds (HVCs). However, in our model, these objects must be (or have been)
associated with DM halos. While Galactic HVCs do appear to require DM
\citep{BW04},
Galactic GCs don't
(e.g. \citep{Sollima+09}), perhaps because they are closer and more tidally
stripped. 

\section*{Acknowledgments}

We thank Joe Silk
 for useful discussions, the Editors for useful suggestions and Jarle
Brinchmann and Anna Gallazzi for respectively supplying us with the stellar masses
and fractions of young galaxies in the SDSS.


\begin{thebibliography}{23}
\providecommand{\natexlab}[1]{#1}
\providecommand{\url}[1]{\texttt{#1}}
\expandafter\ifx\csname urlstyle\endcsname\relax
  \providecommand{\doi}[1]{doi: #1}\else
  \providecommand{\doi}{doi: \begingroup \urlstyle{rm}\Url}\fi

\bibitem[{Aloisi} et~al.(2007){Aloisi}, {Clementini}, {Tosi}, {Annibali},
  {Contreras}, {Fiorentino}, {Mack}, {Marconi}, {Musella}, {Saha}, {Sirianni},
  and {van der Marel}]{Aloisi+07}
A.~{Aloisi}, G.~{Clementini}, M.~{Tosi}, et al.
\newblock  2007, \emph{\apjl}, 667, L151

\bibitem[{Baldry} et~al.(2008){Baldry}, {Glazebrook}, and {Driver}]{BGD08}
I.~K. {Baldry}, K.~{Glazebrook}, and S.~P. {Driver},
\newblock 2008, {\mnras}, 388, 945

\bibitem[{Bell} et~al.(2003){Bell}, {McIntosh}, {Katz}, and
  {Weinberg}]{BMcIKW03}
E.~F. {Bell}, D.~H. {McIntosh}, N.~{Katz}, and M.~D. {Weinberg},
\newblock  2003, \emph{\apjs}, 149, 289

\bibitem[{Birnboim} and {Dekel}(2003)]{BD03}
Y.~{Birnboim} and A.~{Dekel},
\newblock  2003, \emph{\mnras}, 345, 349

\bibitem[{Boselli} et~al.(2008){Boselli}, {Boissier}, {Cortese}, and
  {Gavazzi}]{BBCG08}
A.~{Boselli}, S.~{Boissier}, L.~{Cortese}, and G.~{Gavazzi},
\newblock  2008, \emph{\apj}, 674, 742

\bibitem[{Br{\"u}ns} and {Westmeier}(2004)]{BW04}
C.~{Br{\"u}ns} and T.~{Westmeier},
\newblock  2004, \emph{\aap}, 426, L9

\bibitem[{Cattaneo} et~al.(2011){Cattaneo}, {Mamon}, {Warnick}, and
  {Knebe}]{CMWK10}
A.~{Cattaneo}, G.~A. {Mamon}, K.~{Warnick}, and A.~{Knebe},
\newblock 2011, \emph{\aap}, submitted, arXiv:1002.3257

\bibitem[{De Lucia} and {Blaizot}(2007)]{dLB07}
G.~{De Lucia} and J.~{Blaizot},
\newblock  2007, \emph{\mnras}, 375, 2

\bibitem[{Dekel} and {Silk}(1986)]{DS86}
A.~{Dekel} and J.~{Silk},
\newblock  1986, \emph{\apj}, 303, 39

\bibitem[{Gallazzi} et~al.(2005){Gallazzi}, {Charlot}, {Brinchmann}, {White},
  and {Tremonti}]{Gallazzi+05}
A.~{Gallazzi}, S.~{Charlot}, J.~{Brinchmann}, S.~D.~M. {White}, and C.~A.
  {Tremonti},
\newblock  2005, \emph{\mnras}, 362, 41

\bibitem[{Gallazzi} et~al.(2008){Gallazzi}, {Brinchmann}, {Charlot}, and
  {White}]{GBCW08}
A.~{Gallazzi}, J.~{Brinchmann}, S.~{Charlot}, and S.~D.~M. {White},
\newblock  2008, \emph{\mnras}, 383, 1439

\bibitem[{Gnedin}(2000)]{Gnedin00}
N.~Y. {Gnedin},
\newblock  2000, \emph{\apj}, 542, 535

\bibitem[{Guo} et~al.(2010){Guo}, {White}, {Li}, and
  {Boylan-Kolchin}]{GWLBK10}
Q.~{Guo}, S.~D.~M.~{White}, C.~{Li} and M.~{Boylan-Kolchin},
\newblock 2010, \emph{\mnras}, 404, 1111

\bibitem[{Hunt} et al. (2003)]{HTI03}
L.~K. Hunt, T.~X.~Thuan, and Y.~I.~Izotov,
\newblock 2003, \emph{\apj}, 588, 281

\bibitem[{Izotov} and {Thuan}(2004)]{IT04}
Y.~I. {Izotov} and T.~X. {Thuan},
\newblock  2004, \emph{\apj}, 616, 768

\bibitem[{Jiang} et~al.(2008){Jiang}, {Jing}, {Faltenbacher}, {Lin}, and
  {Li}]{Jiang+08}
C.~Y. {Jiang}, Y.~P. {Jing}, A.~{Faltenbacher}, W.~P. {Lin}, and C.~{Li},
\newblock  2008, \emph{\apj}, 675, 1095

\bibitem[{Kere{\v s}} et~al.(2009){Kere{\v s}}, {Katz}, {Fardal}, {Dav{\'e}},
  and {Weinberg}]{Keres+09a}
D.~{Kere{\v s}}, N.~{Katz}, M.~{Fardal}, R.~{Dav{\'e}}, and D.~H. {Weinberg},
\newblock 2009, \emph{\mnras}, 395, 160

\bibitem[{Knollmann} and {Knebe}(2009)]{KK09}
S.~R. {Knollmann} and A.~{Knebe},
\newblock 2009, \emph{\apjs}, 182, 608

\bibitem[{Lee} et al.(2006)]{Lee+06}
H. Lee, E.~D.~Skillman, E.~D.~Cannon, et al.
\newblock 2006, \emph{\apj}, 647, 970

\bibitem[{Mastropietro} et~al.(2005){Mastropietro}, {Moore}, {Mayer},
  {Debattista}, {Piffaretti}, and {Stadel}]{Mastropietro+05}
C.~{Mastropietro}, B.~{Moore}, L.~{Mayer}, et al.
\newblock  2005, \emph{\mnras}, 364, 607

\bibitem[{Neistein} and {Dekel}(2008)]{ND06}
E.~{Neistein} and A.~{Dekel},
\newblock  2008, \emph{\mnras}, 383, 615

\bibitem[{Papaderos} et al. (2002)]{Papaderos+02}
P.~Papaderos, Y.~I.~Izotov, T.~X.~Thuan, et al.
\newblock 2002, \emph{\aap}, 393, 461

\bibitem[{Salvadori} and {Ferrara}(2009)]{SF09}
S.~{Salvadori} and A.~{Ferrara},
\newblock 2009, \emph{\mnras}, 395, L6

\bibitem[{Sollima} et~al.(2009){Sollima}, {Bellazzini}, {Smart}, {Correnti},
  {Pancino}, {Ferraro}, and {Romano}]{Sollima+09}
A.~{Sollima}, M.~{Bellazzini}, R.~L. {Smart}, M.~{Correnti}, E.~{Pancino},
  F.~R. {Ferraro}, and D.~{Romano},
\newblock 2009, \emph{\mnras}, 396, 2183

\bibitem[{Springel} et~al.(2005){Springel}, {White}, {Jenkins}, {Frenk},
  {Yoshida}, {Gao}, {Navarro}, {Thacker}, {Croton}, {Helly}, {Peacock}, {Cole},
  {Thomas}, {Couchman}, {Evrard}, {Colberg}, and {Pearce}]{Springel+05}
V.~{Springel}, S.~D.~M. {White}, A.~{Jenkins}, et al.
\newblock 2005, \emph{\nat}, 435, 629

\bibitem[{Thoul} and {Weinberg}(1996)]{TW96}
A.~A. {Thoul} and D.~H. {Weinberg},
\newblock 1996, \emph{\apj}, 465, 608

\bibitem[{Thuan} et~al.(2011){Thuan}, {Yakobchuk}, and {Izotov}]{TYI10}
T.~X. {Thuan}, T.~M. {Yakobchuk}, and Y.~I. {Izotov},
\newblock 2011, \emph{\apj}, submitted

\bibitem[{Yang} et~al.(2009){Yang}, {Mo}, and {van den Bosch}]{YMvdB09}
X.~{Yang}, H.~J. {Mo}, and F.~C. {van den Bosch},
\newblock  2009, \emph{\apj}, 695, 900

\end{thebibliography}

\end{document}